\documentstyle[prl,amsfonts,aps]{revtex}

\newcommand{\N}{{\Bbb N}}                          
\newcommand{\Z}{{\Bbb Z}}                          
\newcommand{\no}[1]{: \! #1 \! :}                  
\newcommand{\1}{\openone}                          
\newcommand{\vJ}{\mbox{\boldmath $J$}}             
\newcommand{\vS}{\mbox{\boldmath $S$}}             
\newcommand{\vphi}{\bbox{\phi}}                    
\newcommand{\phdagger}{\mathop{\phantom{\dagger}}} 
\newcommand{\psiop}[1]{\psi^{\phdagger}_{#1}}      
\newcommand{\psidop}[1]{\psi^{\dagger}_{#1}}       
\newcommand{\cop}[1]{c^{\phdagger}_{#1}}           
\newcommand{\cdop}[1]{c^{\dagger}_{#1}}            
\newcommand{\bsigma}[1]{\mbox{\boldmath $\sigma$}^{\phdagger}_{#1}} 

\begin{document}
\title{Multichannel Kondo Effect in an Interacting Electron System:  \\ Exact 
Results for the Low-Temperature Thermodynamics} 

\author{Mats Granath and Henrik Johannesson}

\address{Institute of Theoretical Physics \\ Chalmers University of
Technology and G\"oteborg University \\ S-412 96 G\"oteborg, Sweden}

\maketitle
\begin{abstract}
We study the low-temperature thermodynamics of a spin-$S$ magnetic
impurity coupled to $m \ge 2$  degenerate bands of interacting electrons in
one dimension. By exploiting boundary conformal field theory techniques, we derive exact results 
for the {\em possible} impurity thermal and magnetic response. The leading behavior of the impurity
magnetic susceptibility is shown to be insensitive to the electron-electron interaction. In contrast,
there are two types of scaling behavior of the impurity specific heat consistent with the
symmetries of the problem:  
{\em Either} it remains 
the same as for the ordinary multichannel Kondo problem for 
noninteracting electrons {\em or} it acquires a new leading term governed by 
a critical exponent 
$\alpha_m = (K_{\rho,m}^{-1} - 1)/m$, where $K_{\rho,m} \le 1$ is a generalized 
($m-$channel) Luttinger liquid charge parameter measuring the strength of the
repulsive electron-electron interaction.   
We conjecture that the latter behavior is indeed realized when the impurity is  
exactly screened ($m = 2S$).
\end{abstract}

\newpage 

\section{Introduction}

The violation of Fermi-liquid behavior in the normal state of the 
high-$T_c$ superconductors has turned the study of 
{\em non-Fermi liquid} phenomena into a major theme in condensed matter 
physics \cite{nfl}. Additional motivation 
comes from a growing number of experimental realizations of  
low-dimensional electron structures - such as quasi one-dimensional (1D) 
organic conductors \cite{jerome}, or point contact tunneling in fractional 
quantum Hall devices \cite{milliken} - where electron correlations are 
seen to produce manifest non-Fermi liquid behavior.

Two prominent model problems, serving as  
paradigms for the study of non-Fermi liquids, are the {\em Luttinger liquid} 
\cite{haldane,voit} and the (overscreened) {\em multichannel Kondo 
effect} 
\cite{nozieres,andrei-wiegmann,affleck-ludwig,schlottmann}. ''Luttinger 
liquid'' is the code name for 
the universal low-energy behavior of interacting electrons in 1D, 
whereas the multichannel Kondo model describes a magnetic impurity 
coupled via a spin exchange to 
several degenerate bands of non-interacting electrons in 3D. Both problems have yielded 
to exact solutions, exhibiting a wealth of properties not contained in 
the standard Fermi liquid picture of the metallic state.

Here we consider a spin-$S$ magnetic impurity coupled to $m \ge 2$ 
degenerate bands of {\em interacting} electrons in 1D, thus extending the 
ordinary multichannel Kondo model to the case of interacting electrons.
Specifically, we address the question about the influence of 
electron-electron interaction on the low-temperature thermal and magnetic 
response of the impurity-electron (screening-cloud) composite. This 
problem is particularly interesting as it involves the interplay 
between two kinds of electron correlations; one induced by the 
spin exchange interaction with the impurity, the other coming from the 
direct electron-electron interaction.  
Moreover, the study of a magnetic impurity in the presence of
interacting electrons may shed new light on possible experimental
realizations of the multichannel Kondo effect, such as certain Uranium-containing 
heavy-fermion materials \cite{cox}, or the coupling of conduction electrons
to structural defects in metal point contacts \cite{ralph}.  

We study the problem using  boundary conformal 
field theory (BCFT) \cite{cardy}, assuming that at low temperatures the impurity-electron 
interaction renormalizes onto a scale-invariant boundary condition on the 
bulk theory. This approach, first suggested by Affleck and Ludwig 
for the ordinary multichannel Kondo problem \cite{affleck-ludwig,review}, has  
successfully been employed for a single channel of interacting electrons 
coupled to a spin-1/2 impurity \cite{frojdh,durg}. 
In the present case, with an arbitrary number of electron channels $m \ge 2$, a
BCFT analysis allows for a complete classification of all {\em possible}
critical behaviors of the impurity-electron composite. Being exact, this 
information should prove useful as a guide to - and a test of the validity of - other, 
more direct approaches to the problem (yet to be carried out). 
Specifically, we shall show that 
conformal invariance together with the internal symmetries of the problem 
restrict the possible types of critical behavior to only two: {\em 
Either} the theory is the same as for noninteracting electrons {\em or} 
the electron interaction generates a specific boundary operator that produces a 
new leading behavior of the impurity thermal response. 
In both cases the leading impurity magnetic 
response is insensitive to the electron-electron interaction, implying 
that the screening of the impurity is realized in the same way as in the 
noninteracting problem, with over-, exact, or underscreening depending on 
the number of channels and the magnitude of the impurity spin.  
While our method cannot pinpoint which of the two scenarios is actually realized, we conjecture - 
guided by analogous results for the one-channel problem 
\cite{furusaki,egger} - that electron interactions in the bulk {\em do} induce an anomalous term 
in the
impurity specific heat in the case of exact screening $(m = 2S)$.

\section{The Model}

As microscopic bulk model we take a multiband Hubbard chain with repulsive 
on-site interaction $U>0$, 
\begin{equation}
\label{Hlattice}
H_{el} = -t\! \sum_{n, i, \sigma}( \cdop{n,i\sigma} \cop{n+1,i\sigma} + 
h.c.)
+ U\!\sum_{n,ij,\mu \sigma} \! n_{n,i\sigma}n_{n,j\mu},
\end{equation}
where $\cop{n,i\sigma}$ is the electron operator at site $n$, with $i= 1,....,m$ and  
$\sigma= \uparrow, \downarrow$ band- and spin 
indices, respectively, and $n_{n,i\sigma}=\cdop{n,i\sigma} \cop{n,i\sigma}$ is
the number operator. This model - and its variants - has been extensively studied in the
literature, most recently in \cite{balatsky} where it was argued that 
enhanced superconducting fluctuations may result when the degeneracy of
the on-site interband coupling $U$ is properly lifted.
At large wavelengths we can perform a continuum limit  $\cop{n,i\sigma} \rightarrow 
\sqrt{\frac{a}{2\pi}} \Psi_{i\sigma}(na)$ (with $a$ the lattice spacing), 
which, for small $U$ and away from half-filling, takes (\ref{Hlattice}) onto
\begin{eqnarray}
\label{bulkHamiltonian}
H_{el} = &\frac{1}{2\pi}&\int dx  \biggl\{  v_F
\biggl[\no{ \psidop{L,i\sigma}(x) i \frac{d}{dx} \psiop{L,i\sigma}(x) }
- \no{ \psidop{R,i\sigma}(x) i \frac{d}{dx} \psiop{R,i\sigma}(x) } \biggr]
\nonumber \\
&+&\frac{g}{2}\no{ \psidop{r,i\sigma}(x) \psiop{r,i\sigma}(x) 
\psidop{s, j\mu}(x) \psiop{s, j\mu}(x) } 
+ g\no{\psidop{L,i\sigma}(x) \psiop{R,i\sigma}(x) 
\psidop{R, j\mu}(x) \psiop{L, j\mu}(x) } \biggr\}.
\end{eqnarray}
Here $\psi_{L/R,i\sigma}(x)$ are the left/right moving components
of the electron field $\Psi_{i\sigma}(x)$, expanded about the 
Fermi points $\pm k_F$: $\Psi_{i\sigma}(x)=e^{-ik_Fx}\psiop{L,i\sigma}(x)+
e^{ik_Fx}\psiop{R,i\sigma}(x)$. Summation over repeated indices for band, 
spin, and chirality $r,s=R,L$ is implied. 
The normal ordering is defined w.r.t. the filled Dirac sea, and  
$v_F$ and $g$ are given by $v_F=2at\sin(ak_F)$ and $g=Ua/\pi$ respectively. 

For the purpose of implementing BCFT techniques in the presence of a Kondo impurity
(yet to be added),  we  
decouple charge-, spin-, and band- {\em (flavor-)} degrees of freedom in 
(\ref{bulkHamiltonian}) by a Sugawara construction 
\cite{fuchs}, using the
$U(1)$ (charge),  $SU(2)_m$ (level $m$, spin), and $SU(m)_2$ (level $2$, 
flavor) Kac-Moody currents: 
\begin{eqnarray}
J_{r}(x)&=&\no{ \psidop{r,i\sigma}(x) \psiop{r,i\sigma}(x) }
\label{U1charge} \\ 
\vJ_r(x)&=&\no{\psidop{r,i\sigma}(x)\frac{1}{2}
\bsigma{\sigma\mu} \psiop{r,i\mu}(x)} \label{SU2spin} \\
J_r^A(x)&=&\no{\psidop{r,i\sigma}T_{ij}^A\psiop{r,j\sigma}}, 
\label{SU2flavor}
\end{eqnarray} 
where $\bsigma{}$ are the Pauli matrices, and $T^A, A\!\in\{1,\ldots,m^2-1\}$,
are generators of the defining representation of $SU(m)$ with normalization
$\mbox{tr}T^AT^B = 1/2\delta^{AB}$. 
Diagonalizing the charge sector by a Bogoliubov transformation 
$J_{L/R}=\cosh(\theta) j_{L/R}-\sinh(\theta) j_{R/L}$, with 
$\coth({2\theta})=1+v_F/((2m-1)g)$, we obtain the critical bulk Hamiltonian 
\begin{equation}
\label{sugawara-hamiltonian}
H^*_{el} =\!\frac{1}{2\pi}\int\! dx \biggl\{ \frac{v_c}{4m}
\no{j^i_L(x) j^i_L(x)}
\!+\frac{v_s}{m+2}\no{{\vJ}^i_L(x)\!\cdot\!\vJ^i_L(x)}
+\frac{v_f}{m+2}\no{J^{iA}_L(x) J^{iA}_L(x)}\biggr\},
\end{equation}
where $v_c\!=\!v_F(1+2(2m-1)g/v_F)^{1/2}, v_s\!=\!v_f\!=\!v_F\!-\!g$.
We have here retained only exactly marginal terms in the interaction by removing two
(marginally) irrelevant terms in the spin and flavor sectors \cite{footnote0}.
We have also replaced the right-moving currents with a second species
(labeled by ``2'') of left-moving currents: $j_{L/R}^2(x)\equiv 
j_{R/L}(-x)$ for $x>0$ (with $j_{L/R}^1(x)\equiv j_{L/R}(x)$), and
analogously for spin and flavor currents. This amounts to 
folding the system onto the positive $x$-axis, with a boundary condition 
\begin{equation}
j_L^{1/2}(0)  =  j_R^{2/1}(0) 
\label{trivialBC}
\end{equation}
at the origin, and then analytically continuing the currents back to the full $x-$axis. 
Here (\ref{trivialBC}) simulates the continuity at $x=0$ of the
original bulk theory (with the analogous boundary conditions in 
spin- and flavor sectors). The Sugawara form of $H^*_{el}$ in  
(\ref{sugawara-hamiltonian}) implies invariance 
under independent $U(1)^i$, $SU(2)^i_m$, and $SU(m)^i_2$ transformations (with
$i=1,2$ labeling the two species), reflecting the {\em chiral} symmetry of the 
critical bulk theory.

We now insert a local spin ${\vS}$ at $x=0$, and couple it to the 
electrons by an antiferromagnetic $(\lambda > 0)$ spin exchange interaction
\begin{equation}
\label{HKondo}
H_{\!K}\!=\!\lambda\!\no{(\psidop{L,i\sigma}(0)\! +\! \psidop{R, i\sigma}(0))
\frac{\bsigma{\sigma\mu}}{2}
(\psiop{L,i\mu}(0)\! +\! \psiop{R, i\mu}(0))\! \cdot\! {\vS}}. 
\end {equation}
In the low-energy limit the impurity is expected \cite{review} to renormalize
to a conformally invariant boundary condition on the bulk theory, changing
(\ref{trivialBC}) into a new nontrivial boundary condition on $H^*_{el}$.
By using BCFT to extract the set of {\em boundary operators} present for 
this boundary condition, finite-temperature effects due to the impurity 
can be accessed via standard finite-size scaling by treating (Euclidean) time as 
an inverse temperature. 

\section{Impurity Critical Behavior}
The set of boundary operators ${\cal O}_j$ and the corresponding {\em boundary
scaling dimensions} $\Delta_j$ can be derived from the finite-size spectrum
of $H^*_{el}$ through a conformal mapping from the half-plane to a 
semi-infinite strip. The mapping is such that the boundary condition 
corresponding to the impurity on the half-plane is mapped to both sides of 
the strip, with the boundary operators in the plane in one-to-one 
correspondence to the eigenstates on the strip. In particular, the boundary
dimensions are related to the energy spectrum through the relation
$E=E_0+\pi v \Delta/\ell$, with $E_0$ the groundstate 
energy, and $\ell$ the width of the strip \cite{cardy}.
  
A complete set of eigenstates of $H^*_{el}$ are given by the charge-, spin- and flavor 
{\it conformal towers} , each tower defined  by            
a {\it Kac-Moody primary state} and its descendants \cite{fuchs}. 
The primaries are labeled by $U(1)$ quantum numbers $q^i$ in the
charge sector, $SU(2)$ quantum numbers $j^i$ in the spin sector 
and $SU(m)$ quantum numbers {\em (Dynkin labels)} 
$(m^i_1,\ldots,m^i_{m-1})$ in the flavor 
sector:  
\begin{equation}
q^{{}^1_2} = C^{{}^1_2}\frac{e^{\theta}}{2} \pm  
D^{{}^1_2}\frac{e^{-\theta}}{2}, \ \ \
 j^i=0,\frac{1}{2},...,\frac{m}{2}, \ \ \
\sum_{k=1}^{m-1}m^i_k=0,1,2 \ ,
\label{QuantumNumbers}
\end{equation}
where $C^i, D^i \in \Z$ and $m^i_k\in\N$, with $i=1,2$ labelling the two species as above.
We can express the complete energy spectrum,
and consequently the
complete set of possible boundary scaling dimensions $\Delta=\Delta^1+\Delta^2$ in terms of the
quantum numbers in (\ref{QuantumNumbers}):
\begin{equation}
\Delta^i=\frac{(q^i)^2}{4m}+\frac{j^i(j^i+1)}{m+2}+
\frac{1}{2(m+2)}
\sum_{k=1}^{m-1} \!m^i_k\!\left(\!f_m(k,k)\!+\!\sum_{l=1}^{m-1} m^i_l
\frac{f_m(k,l)}{m}\right)+{\cal N},
\label{Dimensions}
\end{equation}
where $f_m(k,l)\!=\!\mbox{min}(k,l)(m-\mbox{max}(k,l))$ and ${\cal N}\in\N$.

Each conformal boundary condition corresponds to a {\em selection rule} which 
specifies that only certain combinations of conformal towers are allowed. 
Since the trivial boundary condition (\ref{trivialBC}) simply defines the 
bulk theory in terms of a boundary theory, the associated selection rule  
reproduces the {\em bulk} scaling dimensions of 
$H^*_{el}$ \cite{frojdh}. It is less obvious how to identify the correct selection rule 
for the nontrivial boundary condition representing (\ref{HKondo}).  
Fortunately, we do not need the full selection rule to extract the 
leading impurity critical behavior. For this purpose    
it is sufficient to identify the {\em leading 
irrelevant boundary operator} \cite{affleck-ludwig} (LIBO) that can appear in 
the scaling 
Hamiltonian, 
as this is the operator that drives 
the dominant response of the impurity. 
As the possible correction-to-scaling-operators are
boundary operators constrained by the symmetries of the Hamiltonian,  
this sets our strategy: We 
consider {\em all} selection rules for combining conformal towers in 
(\ref{QuantumNumbers}) (thus exhausting all 
conceivable boundary fixed points), for each selecting the corresponding 
LIBO, using (\ref{Dimensions}). We then extract the {\em possible} impurity critical
behaviors by identifying those LIBOs that  (i) {\em produce a  
noninteracting limit $g \rightarrow 0$ consistent with known results, and} (ii)
 {\em respect 
the symmetries of $H^*_{el} + H_K$.}
\subsection{Overscreening: $m > 2S$}

Let us first focus on the case $m > 2S$. Here the noninteracting 
$(g\!=\!0)$ problem renormalizes to a nontrivial fixed point, as 
can be seen by passing to a  basis of definite-parity $(P= \pm)$ fields
$\psiop{\pm,i\sigma}(x) = (1/\sqrt{2})(\psiop{L,i\sigma}(x) \pm
\psiop{R,i\sigma}(-x) )$ \cite{footnote2}. In this basis 
$H_{el}^*[g\!=\!0] + H_K$ 
becomes identical to the Hamiltonian representing 3D noninteracting 
electrons in $2m$ channels $(P=\pm,
i=1,...m)$, coupled to a local spin in the $m$ positive parity channels only. 
At low temperatures this system flows 
to the overscreened $m-$channel Kondo fixed point with a LIBO of 
dimension $\Delta = (4+m)/(2+m)$ 
\cite{pang}. Consider first the case that this fixed point is stable 
against perturbations in $g$ (or connected to a line of $g>0$ boundary fixed points via an
exactly marginal operator). 
To search for a novel {\em leading} scaling 
behavior for $g \neq 0$ it is then sufficient to search 
for boundary operators with dimensions in the interval $1 \le \Delta \le (4+m)/(2+m)$ 
which produce an impurity response analytically connected to that of the 
non-interacting theory.
An operator with $1 \le \Delta < 3/2$ contributes an impurity specific heat 
scaling as $(\Delta-1)^2 T^{2\Delta-2}$ \cite{affleck-ludwig}, 
and condition (i) then 
requires that, as $g \rightarrow 0$, $\Delta_{LIBO} \rightarrow (4+m)/(2+m)$ {\em or} 
that $\Delta_{LIBO} \rightarrow 1$ (with in this case next-leading dimension 
$\Delta \rightarrow (4+m)/(2+m)$). On the other hand, if the $g=0$  
fixed point gets destabilized as $g$ is switched on, condition 
(i) enforces the LIBO at the new $g > 0$ boundary fixed point  
to become marginally relevant for $g=0$ (so as to produce the necessary 
flow back to the known $g=0$ overscreened $m-$channel fixed point) \cite{footnote1}. Thus, 
for this case 
condition (i) unambiguously requires that $\Delta_{LIBO} \rightarrow 1$ as $g \rightarrow 0$.    
  
Turning to condition (ii), we note that the Kondo interaction 
(\ref{HKondo}) couples $L$ and $R$ fields and thus breaks the chiral 
gauge invariance in all three sectors. This implies that the Kac-Moody 
symmetries gets broken down to their diagonal subgroups, i.e.  
$U(1)^1\times U(1)^2 \rightarrow U(1)$ in the charge sector, 
$SU(2)_m^1 \times SU(2)_m^2 \rightarrow SU(2)_{2m}$ in the spin sector, and
$SU(m)_2^1 \times SU(m)_2^2 \rightarrow SU(m)_{4}$ in the flavor sector.  
Operators with
non-zero values of $q_1$ and $q_2$ may thus appear in the charge sector, 
provided that $q_1 = - 
q_2$ as required by conservation of total charge. Similarly, operators in spin- and 
flavor sectors with non-zero quantum numbers $j^i$ and $m^i_j$ 
are now allowed, provided that they  
transform as singlets under the diagonal subgroups.
Remarkably, {\em there exists precisely one generic class of g-dependent 
operators which satisfy condition (i) and (ii).} It is 
given by
\begin{equation}
{\cal O}_1 \sim  \, \no{\mbox{exp}(\frac{i\sqrt{\pi}}{2mK_{\rho,m}} 
\, \phi^1_L)} \times  \no{\mbox{exp}(\frac{i\sqrt{\pi}}{2mK_{\rho,m}} 
\, \phi^2_L)} \times \, \varphi^s \times \varphi^f
\label{LIBO}
\end{equation}
of dimension $\Delta_{{\cal O}_1} = 1+(K^{-1}_{\rho,m}-1)/2m 
\rightarrow 1$ 
as $g \rightarrow 0$. 
Here $\phi^i_L(x)$ is a chiral charge boson of 
species $i$ (i.e. $\phi^i_L(x) = \int dx \ j_L^i(x)$), while
$\varphi^s$ and $\varphi^f$ are the singlet fields (under the diagonal 
subgroups)
in the decomposition of the product of primary fields, 
$(j^1\!=\!1/2) \times (j^2\!=\!1/2)$ and 
$(1,0,...,0) \times (0,0,...,1)$ 
in spin- and flavor sectors, respectively. These carry dimensions 
$\Delta_{\varphi^s} = 3/(2(m+2))$ and $\Delta_{\varphi^f} = 
(m^2-1)/(m(m+2))$.
The parameter $K_{\rho,m}$ is given by
\begin{equation}
K_{\rho,m} 
= (1+2(2m-1)\frac{g}{v_F})^{-\frac{1}{2}} = \frac{v_F}{v_c} \le 1
\label{chargeparameter}
\end{equation}
and plays the role of a generalized (channel-dependent) Luttinger liquid 
parameter \cite{voit}.

The next-leading generic irrelevant boundary operator satisfying (i) and (ii) 
is independent of $g$, and given by
\begin{equation}
{\cal O}_2 
\sim  \vJ^1_{-1} \cdot  
\vphi^1\times\1^2 + \1^1\times\vJ^2_{-1} \cdot \vphi^2 ,
\label{NextLIBO}
\end{equation}
with $\vJ^i_{-1} \cdot \vphi^i$ 
the first Kac-Moody descendant of the spin-1 primary field 
$\vphi^i$, obtained by contraction with the Fourier mode $\vJ^i_{-1}$ of the  
$SU(2)_m$ currents. It
carries dimension $\Delta_{{\cal O}_2} = (m+4)/(m+2)$, and is the same 
operator that drives the leading impurity response in the non-interacting 
problem. For certain special values of $m$ additional $g$-dependent boundary operators 
satisfying (i) and (ii) appear, but as these are non-generic and, for 
given $m$, of higher dimensions than $\Delta_{{\cal O}_1}$, we do not consider 
them here \cite{NonGeneric}.

Piecing together the results, it follows 
that either ${\cal O}_1$ in (\ref{LIBO}) {\em 
or} ${\cal O}_2$ in (\ref{NextLIBO}) plays the role of a LIBO at the $g>0$ boundary fixed 
point. 
We may thus define a scaling Hamiltonian 
\begin{equation}
H_{scaling} = H^*_{el} + \mu_1 {\cal O}_1(0) + \mu_2 {\cal 
O}_2(0) + \mu_3 {\cal O}_3(0)..., 
\label{ScalingHamiltonian}
\end{equation}
with $\mu_j$ conjugate scaling fields, and ${\cal O}_{j>2}$ less relevant 
operators. In the case that ${\cal O}_1$ in (\ref{LIBO}) does {\em not} 
appear, $\mu_1 \equiv 0$. Using 
(\ref{ScalingHamiltonian}), the thermal response may now 
be calculated perturbatively 
in the scaling fields $\mu_j$, using standard techniques 
\cite{affleck-ludwig}. We thus obtain for the impurity specific heat:
\begin{eqnarray}
C_{imp} & = & c_1(1-K_{\rho,m}^{-1})^2T^{(K^{-1}_{\rho,m} - 1)/m} 
 +  \left\{ \begin{array}{ll} 
     c_2 T \mbox{ln}(\frac{T_K}{T}) + ... & m=2, S= \frac{1}{2} \nonumber \\
        c_2' T^{\frac{4}{m+2}} + ... & m > 2, m > 2S \nonumber \\
            \end{array} \right .  \ \ T \rightarrow 0. \\  
\label{scaling}
\end{eqnarray}
Here $c_1, c_2$ and $c_2'$ are amplitudes of second order in the 
scaling fields with corresponding indices, $T_K$ plays the role of a Kondo temperature,
 and ``...'' denotes subleading terms.   
Note that the amplitude of the leading term in (\ref{scaling}) always vanishes when $g=0$, 
thus making the second $g-$independent term dominant. 

The result in (\ref{scaling}) is exact and independent
of the precise nature of the boundary fixed point. In the case that $\mu_1 \equiv 0$,
and hence $c_1 \equiv 0$, the
$g > 0$ fixed point is the {\em same} as for the ordinary (noninteracting) overscreened
problem, although the content of subleading irrelevant operators ${\cal O}_{j>2}$ may
differ. In the alternative case, with $\mu_1 \neq 0 \ (c_1 \neq 0)$, the situation is more intricate,
with, in principle, three possibilities: (a) the $g=0$ (ordinary overscreened Kondo) and
$g \neq 0$ fixed points are the same, but with different contents of irrelevant
operators, (b) the $g=0$ and $g > 0$ fixed points are continuosly 
connected via a critical line by an exactly
marginal boundary operator with a scaling field parameterized by $g$, or (c) the $g=0$
and $g>0$ fixed points are distinct and the flow between them is governed by a marginally
relevant operator \cite{footnote1}.
We postpone a discussion of the various possibilities to the next section. 

Turning to the impurity magnetic susceptibility $\chi_{imp}$, its leading scaling 
behavior is produced by the lowest-dimension boundary operator which 
contains a {\em nontrivial} singlet $SU(2)_{2m}$ factor that couples to the 
total spin density \cite{frojdh}. 
Since ${\cal O}_1$ in (\ref{LIBO}) only contains the 
identity in the $SU(2)_{2m}$ sector the desired operator is identified as ${\cal O}_2$ in 
(\ref{NextLIBO}). Thus, the {\em leading} term in the magnetic susceptibility 
$\chi_{imp}$ due to the impurity 
is independent of the electron-electron interaction, and remains the 
{\em same} as for the noninteracting overscreened $m-$channel Kondo problem 
\cite{andrei-wiegmann,affleck-ludwig}:
\begin{eqnarray}
\chi_{imp}& = & \left\{ \begin{array}{ll} {c}_2\ln({\frac{T_K}{T}}) +...  
& m=2, S=1/2 \\ 
		{c'}_2 T^{\frac{2-m}{m+2}}+ ... & m>2, m>2S \end{array} \right . \ \ \ \ T 
		\rightarrow 0 
\label{impsusc}
\end{eqnarray}
where $c_2$ and ${c'}_2$ are second order in scaling fields
and "..." denotes subleading terms. We find that there are no subleading 
interaction-dependent divergent contributions possible, as these would give rise to 
new divergences also in the noninteracting limit. 

\subsection{Exact screening and underscreening: $m \le 2S$} 

An analysis analogous to the one above can be carried out for $m \le 
2S$ as well. Again passing to a definite-parity basis and exploring known results 
\cite{schlottmann}, one verifies that the {\em noninteracting} 1D electron groundstate 
carries a spin $S-m/2$ corresponding to a strong coupling fixed point. When $m=2S$ the
impurity is completely screened and the situation is essentially the same as for the
ordinary single-channel Kondo problem with impurity spin 1/2: the electron screening cloud
behaves as a local Fermi liquid with a $\pi/2$ phase shift of the single-electron wave
functions. Analogous to the single-channel problem \cite{frojdh}, there are three degenerate
LIBOs for this case, given by the energy-momentum tensors (of dimension $\Delta=2$) in
charge-, spin-, and flavor sectors. When $m=2S$ these produce the {\em leading} term in the 
impurity specific heat, $C_{imp} = b_1T + ... $,
as well as in the susceptibility, $\chi_{imp} = b_2 - b_3T^2 + ....$, 
with $b_{1,2,3} > 0$ amplitudes linear in the scaling fields, and with higher powers
in temperature coming from higher-order descendants of the identity operator. When $m<2S$ the
impurity spin is only partially screened as there are not enough conduction electron
channels to yield a singlet groundstate. This leaves an asymptotically 
decoupled spin $S-m/2 >
0$, adding a Curie-like contribution to $C_{imp}$ and $\chi_{imp}$, 
in addition to logarithmic corrections characteristic of
asymptotic freedom \cite{schlottmann}.

Let us study the case $m=2S$ and explore what happens when turning on the electron
interaction. Implementing condition (i) from the previous section, possible LIBOs appearing
for $g>0$ must have dimensions $\Delta$ with the property $\Delta \rightarrow 1$ or $\Delta
\rightarrow 2$ as $g \rightarrow 0$. Using condition (ii), we find that in addition
to ${\cal O}_1$ in (\ref{LIBO}) there are several new allowed classes of $g-$dependent 
generic boundary operators, all
with $\Delta \rightarrow 2$ as $g \rightarrow 0$. As any boundary operator with dimension
$\Delta > 3/2$ produces the same {\em leading} scaling in temperature as a $\Delta =2$ operator
(although of different amplitudes) \cite{frojdh}, the only {\em possible} leading  
term with an interaction-dependent exponent is again generated
by ${\cal O}_1$, as in the overscreened case. Thus, since ${\cal O}_1$ does not contribute to the
impurity susceptibility, its leading behavior
remains the same as in the non-interacting problem, exhibiting exact
screening with a constant zero-temperature contribution,
\begin{equation}
 \chi_{imp} = b_2 + ... - b_3T^2 + ....\ , \ \ \ m=2S, \ g \ge 0, \ T \rightarrow 0
\label{gspecific}
\end{equation}
where ``....'' denotes possible second-order contributions in scaling 
fields. The amplitude $b_2$ is the same as in the noninteracting problem, 
while $b_3$ may pick up second-order interaction-dependent terms contributed by 
subleading operators. 
The leading {\em possible} interaction-dependent correction to (\ref{gspecific}) scales as
\begin{equation}
\chi_{imp}^{corr} \sim(K^{-1}_{\rho,m}-1)T^{1+(K^{-1}_{\rho,m}-1)/m}
\label{corr}
\end{equation}
and is produced at second order by the composite boundary operator
\begin{equation}
{\cal O}_3 \sim \no{\mbox{exp}(\frac{i\sqrt{\pi}}{2mK_{\rho,m}}
\, \phi^1_L)} \times  \no{\mbox{exp}(\frac{i\sqrt{\pi}}{2mK_{\rho,m}}
\, \phi^2_L)} \times \, \vJ^{diag}_{-1}\cdot\mbox{\boldmath $\varphi$}^s 
\times \varphi^f.
\label{corrop}
\end{equation}
Here $\vJ^{diag}\equiv \vJ^1_L+\vJ^2_L$ is the generator of the diagonal
$SU(2)_{2m}$ subgroup in the spin sector, $\mbox{\boldmath $\varphi$}^s$ is a diagonal
spin-$1$ field in the product of primaries  
$(j^1\!=\!1/2) \times (j^2\!=\!1/2)$,and the charge and flavor
factors are the same as for ${\cal O}_1$ above. The operator ${\cal O}_3$ has scaling dimension
$\Delta_3=1+\Delta_1=2+(K^{-1}_{\rho,m}-1)/2m$ and, as seen in 
(\ref{corr}), gives a vanishing amplitude at $g=0$, thus ensuring the 
correct behavior in the noninteracting limit.
Whether ${\cal O}_3$ appears in the spectrum or not, however, must be
checked by an independent method.

Two possible 
scenarios again emerge for the scaling of the impurity specific heat: {\em Either} it 
remains the same as in the noninteracting exactly screened problem 
{\em or}, in the case that ${\cal O}_1$ appears as a LIBO: 
\begin{equation}
C_{imp} = c_1(1-K^{-1}_{\rho,m})^2 T^{(K^{-1}_{\rho,m} - 1)/m} 
+ b_1T + ... \ , \ \ \ m=2S, \ g \ge 0, \ T \rightarrow 0 
\label{NFLS}
\end{equation} 
where the amplitude $c_1$ is independent of $K_{\rho, m}$, and $b_1$ 
may differ from the noninteracting problem by second order additive terms coming 
from interaction-dependent subleading corrections. Notably, by putting 
$m=1$ in (\ref{NFLS}) 
we recover the critical exponent conjectured by Furusaki and Nagaosa
\cite{furusaki} for the impurity specific heat in the (exactly screened) single-channel problem.
The leading {\em possible} interaction-dependent correction to (\ref{NFLS}) is also produced
by ${\cal O}_3$ and scales as 
\begin{equation}
C_{imp}^{corr} \sim (K_{\rho,m}^{-1} -1) T^{2+(K^{-1}_{\rho,m}-1)/m},
\label{spec-corr}
\end{equation}
with a vanishing amplitude in the noninteracting limit.

For $m<2S$, the influence of the electron-electron interactions on the 
impurity-electron composite corresponding to the screened part of the 
impurity spin is the same as for exact screening, with the same
scenarios for the critical behavior. However, the weak coupling between the
uncompensated (asymptotically free) part of the impurity spin and the conduction 
electrons produce corrections at finite $T$ \cite{schlottmann} that may get modified by 
the electron-electron interaction. We have not attempted to include these effects here.

\section{Discussion} 

To conclude, we have presented an analysis of the possible
low-temperature thermodynamics of the multichannel Kondo problem for an
interacting electron system in 1D. While the leading term in the impurity susceptibility
remains the same
as for noninteracting electrons (for any number of channels $m$ and
impurity spin $S$), there are two possible behaviors for the impurity specific
heat consistent with the symmetries of the problem: {\em Either} it
remains the same as in
the noninteracting problem {\em or} it acquires a new leading term,
scaling with a non-Fermi liquid exponent $\alpha_m =
(K_{\rho,m}^{-1} - 1)/m$, with $K_{\rho,m} \le 1$ in (\ref{chargeparameter})
measuring the strength of the repulsive electron-electron interaction.
These results are exact, given the existence of a stable
boundary fixed point, an assumption common to all applications of BCFT to 
a quantum impurity
problem \cite{review}.
As we discussed in Sec. III. A,
this fixed point (for given $m$ and $S$) is disconnected from that of the
noninteracting
problem only if the LIBO ${\cal O}_1$ in (\ref{LIBO}) turns marginally relevant as $g$ is
put to zero. Does this happen? A conclusive answer would require a construction of the
corresponding exact renormalization group equations. This is a nontrivial 
task, and we have only carried out a perturbative analysis to second 
order in the scaling fields. Turning (\ref{ScalingHamiltonian}) into a Lagrangian, and
integrating out the short-time degrees of freedom using the operator product
expansion \cite{CardyBook}, one obtains the 1-loop RG equation for the scaling field $\mu_1$
conjugate to ${\cal O}_1$:  
\begin{equation}
\frac{d\mu_1}{d\ln{\tau_0}} = -(q_1+q_2)(\lambda_1+\lambda_2)\mu_1 \ ,
\end{equation}
with $\tau_0$ a short-time cut-off,  
$q_1$ and $q_2$ the charge quantum numbers of ${\cal O}_1$, and
$\lambda_1$ and $\lambda_2$ the scaling fields of 
the exactly marginal charge currents
$j^1_L$ and $j^2_L$ (allowed due to breaking of particle-hole symmetry 
in the microscopic Hamiltonian (\ref{Hlattice}) \cite{Cbreaking}). Conservation of total charge 
$q_1+q_2=0$ thus gives  $d\mu_1/d\ln{\tau_0}=0$ to second order in the 
scaling fields. 
If this property does persist to higher orders (as suggested by the cancellation of the
1-loop contribution due to a symmetry), the $g\neq0$ and $g=0$ boundary fixed points
are either a) the same (but with different contents of irrelevant operators) {\em or} b)
connected via a line of fixed points by the exactly marginal charge currents $j_L^1$ and $j_L^2$
(with scaling fields $\lambda_1$ and $\lambda_2$ parameterized by $g$). As we are unable to
determine the actual scenario, 
the question about the
relation of the $g>0$ fixed point to that of the noninteracting problem remains open.

A second open issue is whether the electron-electron interaction
influences the impurity-electron (screening cloud) composite
differently when the impurity is overscreened $(m >2S)$ as compared to
under- or exact screening $(m \le 2S)$. In the overscreened case with
free electrons the
impurity induces a critical behavior where the size of the screening cloud
diverges as one approaches zero temperature. As a consequence, {\em all}
conduction electrons become correlated due to the presence of
the impurity. By turning on a weak (screened) Coulomb interaction among the electrons
(in 1D simulated by the local e-e interaction in (\ref{bulkHamiltonian}))
these correlations may change. Will the change be such as
to produce a
novel impurity critical behavior? While our analysis does not
provide an answer, it predicts its exact form if it does appear.
In the case of exact screening there is strong evidence
that the impurity scaling behavior is indeed governed by interaction-dependent
exponents. In this case the screening cloud has a finite
extent. Turning on the Coulomb interaction, electrons "outside" of the cloud will
become correlated, most likely influencing the rate
with which they tunnel into and out of the cloud, hence influencing its
properties. As this impurity-electron composite is
described by a Fermi-liquid fixed point (where the electrons simply acquire a
phase shift) in exact analogy with the single-channel Kondo problem, we
expect that the effect of turning on electron interactions will indeed be
similar to the single-channel case. Considering the recent Monte Carlo
data by Egger and Komnik \cite{egger} supporting the single-channel
Furusaki-Nagaosa scaling \cite{furusaki}, this strongly favors the
appearance of the interaction-dependent exponents in (\ref{corr}),
(\ref{NFLS}) and (\ref{spec-corr}) when $m = 2S$.

\subsection*{Acknowledgment}

We thank I. Affleck, N. Andrei, R. Egger, P. Fr\"ojdh, and A. W. W. Ludwig   
for valuable input.
H. J. acknowledges support from the Swedish Natural Science Research Council. 




\begin{references}
\bibitem{footnote2} Note that the interacting part of the bulk Hamiltonian 
$H_{el}$ becomes
nonlocal in the definite-parity basis, thus precluding its use when $g \neq 0$.
\bibitem{NonGeneric} A computer check of the spectra with $2 \le m \le 50$, using 
(\ref{Dimensions}), reveals that additional operators satisfying (i) and 
(ii) occur at $m=2,4,6,8,9,18,32$ and $50$. As there is no change in the structure of the Hamiltonian
for these particular channel numbers, we expect these operators to be excluded by the selection rule
defining the boundary fixed point.
\end{references}
\end{document}